\documentstyle[prc,aps,epsf,epsfig,floats,twocolumn,amsmath,amsfonts]{revtex} 
\newcommand{\beq}{\begin{equation}} 
\newcommand{\eeq}{\end{equation}} 
\newcommand{\be}{\begin{equation}} 
\newcommand{\ee}{\end{equation}} 
\newcommand{\bea}{\begin{eqnarray}} 
\newcommand{\eea}{\end{eqnarray}}

\newcommand{\no}{\nonumber} 
\renewcommand{\arraystretch}{1.1}

\newcommand{\lanl}{\langle} 
\newcommand{\ranl}{\rangle}

\newcommand{\beqa}{\begin{eqnarray}} 
\newcommand{\eeqa}{\end{eqnarray}} 
\newcommand{\beqan}{\begin{eqnarray*}} 
\newcommand{\eeqan}{\end{eqnarray*}} 
\newcommand{\ba}{\begin{array}} 
\newcommand{\ea}{\end{array}}

\newcommand{\del}{\partial}

\newcommand{\bdm}{\begin{displaymath}} 
\newcommand{\edm}{\end{displaymath}}

\def\w{\omega}

\newcommand{\chpt}{$\chi$PT~} 
\begin{document} 
\allowdisplaybreaks{ 
\draft 
\twocolumn[\hsize\textwidth\columnwidth\hsize\csname 
@twocolumnfalse\endcsname 
 
\title{ 
\hfill{\small { FZJ-IKP(TH)-2000-20}}\\[0.2cm] 
Towards an understanding of isospin violation in pion--nucleon scattering}

\author{ 
Nadia Fettes, Ulf-G. Mei{\ss}ner}

\address{Forschungszentrum J\"ulich, Institut f\"ur Kernphysik (Theorie), D-52425  
 J\"ulich, Germany\\ E-Mail addresses: n.fettes@fz-juelich.de, u.meissner@fz-juelich.de } 
\maketitle

\begin{abstract} 
We investigate isospin breaking in low--energy pion--nucleon scattering 
in the framework of chiral perturbation theory. This work extends the 
systematic analysis of~\cite{fmsi} to the energy range above threshold. 
Various relations, which identically vanish in the limit 
of isospin symmetry, are used to quantify isospin breaking 
effects. We study the energy dependence of the S-- and P--wave  
projections of these ratios and find dramatic effects 
in the S--waves of those two relations which are given 
in terms of isoscalar quantities only. This effect drops rather quickly 
with growing  center--of--mass energy. 
 
\end{abstract} 
\medskip 
 
\vspace{1cm} 
 
]  
 
\section{Introduction} 
\label{sec:noiso_intro}} 
 
Pion--nucleon scattering ($\pi N$) is one of the prime reactions 
to test our understanding not only of the spontaneous and explicit chiral symmetry 
breaking QCD is supposed to undergo, but also of isospin symmetry violation.  
The pion--nucleon system is particularly well suited for such an analysis, since  
chiral symmetry breaking and isospin breaking appear at the same 
chiral order.  
For neutral--pion scattering off nucleons, the isospin violating effects  
can be dramatically enhanced due to the smallness of the isoscalar  
pion--nucleon amplitude~\cite{wein,ms}. 
This spectacular effect in the difference of the $\pi^0 p$ and $\pi^0 n$ scattering  
lengths is, however, at present not amenable to a direct experimental  
verification.
It is therefore mandatory to include also the channels with charged pions in any 
analysis of isospin violation.  
The immense experimental effort in the domain of low--energy pion--nucleon physics 
has stimulated considerable interest in using $\pi N$ scattering 
data to extract information about the violation of isospin  
symmetry of the strong interactions~\cite{weinmit,bira}. 
Some analyses indicate effects as large as 7\%~\cite{gibbs,matsi}.  
In both these analyses, the source of this rather large effect remains 
mysterious.  
Since the two methods are independent and based on different approaches, 
it became a challenge to find a theoretical explanation 
for these rather phenomenological observations.  
Microscopically, there are two competing sources of isospin violation,  
which are generally of the same size, namely the strong effect due to  
the light quark mass difference $m_d - m_u \simeq m_u$ and the  
electromagnetic one caused by virtual photons. There is thus need 
to describe both effects, the strong and the electromagnetic interaction, 
consistently. One of the major criticisms raised against 
the analyses of isospin violation performed up to now is their inherent   
incompatibility of electromagnetic and strong effects. 
Chiral perturbation theory ($\chi$PT) is able to remedy this problem. 
Ongoing effort in this field has resulted in the 
development of an effective field theory of pions, nucleons, and virtual photons, 
which now allows for the separation of both sources of isospin violation 
in a consistent fashion. The corresponding effective Lagrangian was developed  
in~\cite{mms,ms,MM} extending the standard $\pi N$ EFT.  
The pertinent power counting of the EFT is based on the (phenomenological)  
observation that besides the pion mass and momenta, the electric charge $e$ 
should be counted as an additional small parameter, given the fact   
that  
\begin{equation} 
{e^2 \over 4\pi} \simeq {M_\pi^2 \over (4\pi F_\pi)^2} \simeq {1 \over 
  100}~. 
\label{photopc} 
\end{equation}  
Note, however, that $M_\pi$ vanishes in the chiral limit, whereas the 
left--hand side of eq.~(\ref{photopc}) remains finite. 
If this counting scheme is applied, we only have to deal with one 
expansion parameter $q$, which corresponds to small momenta, 
pion masses, as well as elementary charges. Since every emitted 
virtual photon will have to be reabsorbed, we will only have 
to consider terms proportional to the square of the charge matrix. 
The difference of the squares of the charged-- and neutral--pion masses  
is thus an effect of second chiral order, i.e.\ a leading order effect. 
On the other hand, the mass difference for the nucleons also starts out at  
second order and is thus suppressed by two chiral orders. 
 
\medskip\noindent 
The analysis of isospin violation in $\pi N$ scattering 
proceeds essentially in three steps. First, one ignores all 
isospin breaking effects, i.e.\ one sets $e= 0$ and $m_u = m_d$.   
This is the approximation on which the analysis in  
refs.~\cite{FMS,pin4} was based. These papers comprise the most 
detailed studies of pion--nucleon scattering in the framework of 
heavy--baryon chiral perturbation theory (related studies have been  
published in~\cite{bkmpin,bkmpin2,bkmci,moj,paul}). 
It is obvious that one needs a precise description of 
the large isospin symmetric ``background'' of the scattering amplitude 
in order to be able to pin down the small isospin violating effects. 
The quality of the results obtained in refs.~\cite{FMS,pin4}  
makes us feel confident that we have a sufficiently accurate starting point. 
 
\medskip\noindent 
In the second step, one should include the leading isospin breaking terms encoded in the 
pion and nucleon mass differences. The mass splitting for the nucleons 
amounts to about 1~MeV, whereas the charged-- to neutral--pion mass difference 
is of the order of 5~MeV. To the accuracy we are working (the third 
order in small momenta 
and charges) one has to consider such effects. The strength of \chpt now lies in 
the fact that one can consistently take into account only the effect from 
those isospin violating low--energy constants which enter the particles' masses. 
This is the approximation which we will consider here. In fact, in  
neutral--pion photoproduction off nucleons, to third order in small momenta, this 
approximation leads to the only isospin breaking effect, which reveals itself 
in the large cusp effect at the secondary threshold (i.e.\ at the $\pi^+ n$ 
threshold in the case of $\gamma p \to \pi^0 p$)~\cite{bkmzpc}.  
 
\medskip\noindent 
In the third step, which goes beyond the scope of this work, one has to account for all  
virtual--photon effects, in particular soft--photon emission from charged--particle legs 
and the Coulomb poles due to the ladder exchange of (soft) virtual photons between 
charged external particles.  
 
\medskip\noindent 
In the last two steps, the notion of partial waves with defined total isospin 
becomes doubtful. It is thus better not to give predictions for any specific 
isospin channel, but rather to consider quantities for measurable pion--nucleon 
reactions. After the third step will have been performed, one will be 
able to directly fit to experimental cross section and polarization data. 
This will have the advantage that one will not have to rely on any 
code for electromagnetic corrections, which might or might not 
be compatible with the hadronic analysis. Instead, one will rather  
be able to compute electromagnetic corrections consistently  
in the framework of chiral perturbation theory. 
Fitting the low--energy constants (LECs) of the full amplitude 
to experimental data will eventually allow to pin down the values 
of the hadronic LECs and to give an unambiguous definition of 
the hadronic phase shifts. 
 
\medskip\noindent 
However, this work will only proceed up to the second step, the inclusion of 
mass difference effects and the leading strong isospin breaking vertices.  
We believe that the essential effects 
of isospin violation are captured in the calculation presented here.  
In~\cite{fmsi}, the size of isospin violation in threshold $\pi N$  
scattering has been investigated. We now extend this analysis  
to the low--energy region above threshold; as in~\cite{fmsi}, 
we will quantify isospin breaking effects by using a set of relations,  
which are fulfilled in the limit of exact isospin symmetry. 
 
\medskip\noindent 
The manuscript is organized as follows. In
section~\ref{sec:noiso_quant} we show how to generalize the
pion--nucleon scattering amplitude in case of isospin violation,
and discuss how to quantify isospin breaking in terms of
ratios, that vanish in the limit of exact isospin symmetry.
Section~\ref{sec:noiso_amplitude} is concerned with 
the calculation of the pertinent $\pi N$ scattering amplitude in
the presence of strong and electromagnetic isospin breaking.
We give results in section~\ref{sec:noiso_results} and compare our approach
to previous analyses of this topic.
We conclude with a short summary and
outlook in section~\ref{sec:noiso_summary}.

\bigskip 
\section{Quantification of isospin symmetry breaking} 
\label{sec:noiso_quant} 
 
\subsection{Generalization of the scattering amplitude 
\label{sec:noiso_genamp}} 
 
We consider the elastic scattering process 
\begin{equation} 
\pi^a (q) + N(k_1) \to \pi^b (q_2) + N(k_2)~, 
\end{equation} 
where $\pi^c (q_i)$ denotes a pion state in the cartesian isospin basis  
with four--momentum $q_i$  
and $N (k_i)$ a proton or a neutron with four--momentum $k_i$.  
The masses of the in-- (out--) going nucleon and pion are 
denoted by $m_{1(2)}$ and  $M_{a(b)}$, respectively  
($M_1 = M_2 = M_{\pi^+},M_3=M_{\pi^0}$). 
Consider the center--of--mass system (cms)  with  
${\vec k}_i= -{\vec q}_i~(i=1,2)$.  
For the nucleons in the heavy--fermion approach, we set  
$k_\mu = m \, v_\mu + p_\mu$ with $v\cdot p \ll m$, and 
$m$ the nucleon mass in the chiral limit.  
The small residual momentum $p_\mu$ is thus a quantity of chiral order one. 
The pion and the nucleon energy in the in--state are (we use $v =(1,0,0,0)$)  
\beq 
\omega_1 = v \cdot q_1 = \sqrt{M_a^2 + {\vec{q}\,}_1^2} \,\, ,  
\quad E_1 = \sqrt{m_1^2 + \omega_1^2 - M_a^2}~, 
\end{equation} 
in order. The energies of the out--going pion and nucleon read 
\beq 
\omega_2  = {(E_1 + \omega_1 )^2 + M_b^2 - m_2^2 \over 2 (E_1 + \omega_1 )} \,\, ,  
\quad E_2 = E_1 + \omega_1 - \omega_2 
\,\,\, , 
\end{equation} 
respectively. We come back to these kinematical relations after 
introducing the pion--nucleon scattering amplitudes appropriate 
for the discussion of isospin violation. 
 
\medskip\noindent 
In the limit of isospin symmetry, 
pion--nucleon scattering can be fully described in terms of two amplitudes  
$T^{1/2}$ and $T^{3/2}$, or equivalently $T^{+}$ and $T^-$. These 
amplitudes depend on two kinematical variables, which we can choose to be 
the pion cms energy $\omega$ and the invariant momentum 
transfer squared $t = (q_1-q_2)^2$. 
In that case, one cannot account  
for any difference in the scattering off protons compared to the one off neutrons. 
Indeed, there is no isospin operator which acts selectively on the nucleons. 
In the presence of isospin violation, i.e.\ isovector symmetry breaking terms such as  
$(m_u -m_d)(\bar{u}u-\bar{d}d)$, one thus has to generalize the 
standard form of the $\pi N$ scattering amplitude to  
\bea  
T^{ba} (\omega , t) &=&   
\delta^{ab} T^+_{ba}  (\omega , t) 
+ \delta^{ab} \tau^3 T^{3+}_{ba}  (\omega , t) \nonumber \\ 
&+& i \epsilon^{bac} \tau^c T^-_{ba} (\omega , t) 
+ i\epsilon^{bac} \tau^c \tau^3 T^{3-}_{ba} (\omega , t)   
\quad , 
\label{noiso_amp1}  
\eea  
in terms of {\it two isoscalar} ($T^{+,3+}_{ba}$) and {\it two isovector}  
($T^{-,3-}_{ba}$) amplitudes.  
It is important to realize that the $T_{ba}^{\pm}$ are exclusively 
sensitive to the neutral-- to charged--pion mass difference. 
The $T_{ba}^{3\pm}$ on the other hand distinguish between the different pion states, 
as well as between scattering off proton or neutron. Indeed the matrix $\tau^3$ 
acts differently on proton and neutron. Therefore, these amplitudes are also sensitive 
to the proton--neutron mass difference. 
The amplitudes are functions of two variables. As in the isospin symmetric case,  
we choose the mean pion energy $\omega$ and the invariant  
momentum transfer squared $t$. 
Since the total mass in the initial state is in general different from 
the mass of the outgoing particles, the energies of the pions 
in the in-- and out--state, $\omega_{1,2}$, will also be no longer equal, 
\bea\label{Dom} 
\Delta \omega  &=& \omega_2 - \omega_1 
 =   \frac{(M_b^2-M_a^2) - (m_2^2-m_1^2)}{2 \sqrt{s}}\no\\ 
&=&  \frac{(M_b^2-M_a^2) - (m_2^2-m_1^2)}{2 m_1}  
\left[1 - \frac{\w_1}{m_1} + {\cal O}(q^2) \right]~, 
\eea 
with $\sqrt{s}$ the total cms energy.  
{}From eq.~(\ref{Dom}) one can easily see that while the pion  
energies $\omega_{1,2}$ are of order $q$,  
their difference only starts out at second order in the chiral 
expansion. This has important 
consequences as will be discussed later. 
The $T$--amplitudes split into a spin--non--flip 
and a spin--flip term, denoted by $g$ and $h$, respectively.  
The most general $\pi N$ scattering amplitude which will allow us to describe  
isospin symmetry violating effects thus reads: 
\bea\label{noiso_amp2} 
T^{ba}_{\pi N}= {\cal N}_1 {\cal N}_2 & \biggl\{& 
\delta^{ab}\, g_{ba}^+ + \delta^{ab}\,\tau^3 \,g_{ba}^{3+}\no\\ 
&+&  i\, \epsilon^{bac} \, \tau^c \, g_{ba}^-  
+  i\, \epsilon^{bac} \, \tau^c\, \tau^3 \, g_{ba}^{3-}\no\\ 
&+& i \vec\sigma \cdot(\vec{q}_2\times \vec{q}_1\,)  
\Big[ 
\delta^{ab}\, h_{ba}^+ + \delta^{ab}\,\tau^3 \,h_{ba}^{3+}\no\\ 
&+&  i\, \epsilon^{bac} \, \tau^c \, h_{ba}^-  
+  i\, \epsilon^{bac} \, \tau^c\, \tau^3 \, h_{ba}^{3-} 
\Big] 
~\biggr\},\no\\ 
\eea 
with ${\cal N}_i =\sqrt{(E_i + m_i)/(2m_i)}\,\, (i=1,2)$ the standard spinor 
normalization~\cite{zfac}. 
We refrain from projecting these amplitudes onto states of defined total 
isospin, since isospin is no good quantum number anymore. But we can still define 
states with given angular and total momentum by 
\bea 
&&f_{l\pm, ba}^{(\pm/ 3\pm)} (s) = {{\sqrt{E_1+m_1} \sqrt{E_2+m_2}} \over 16 \pi \sqrt{s}} \times 
\no\\&&  
\int_{-1}^{+1} dz \, \biggl[\,g^{(\pm/ 3\pm)}_{ba} \, P_l (z)  
+ \vec{q}_1\cdot \vec{q}_2 \, h^{(\pm/ 3\pm)}_{ba} \, \tilde{P}_{l} (z) \biggr]~, 
\label{partwavv} 
\eea 
with $\tilde{P}_l(z) = P_{l\pm1}(z) -zP_l (z)$ a combination of  
the usual Legendre polynomials. 
Isospin indices will now be combined in such a way as to match the 
physical reactions. 
Since the pions come in three and the nucleons in two charge states, we have 
ten reaction channels. In terms of the just defined amplitudes, these 
read: 
\bea 
 f_{\pi^+ p \to \pi^+ p} & = & f_{11}^+ - f_{12}^- + f_{11}^{3+} - f_{12}^{3-}~, \label{noiso_reac1}\\ 
 f_{\pi^- p \to \pi^- p} & = & f_{11}^+ + f_{12}^- + f_{11}^{3+} + f_{12}^{3-}~, \label{noiso_reac2}\\ 
 f_{\pi^0 p \to \pi^0 p} & = & f_{33}^+ + f_{33}^{3+}~,  \label{noiso_reac3}\\ 
 f_{\pi^+ n \to \pi^+ n} & = & f_{11}^+ + f_{12}^- - f_{11}^{3+} - f_{12}^{3-}~, \label{noiso_reac4}\\ 
 f_{\pi^- n \to \pi^- n} & = & f_{11}^+ - f_{12}^- - f_{11}^{3+} + f_{12}^{3-}~, \label{noiso_reac5}\\ 
 f_{\pi^0 n \to \pi^0 n} & = & f_{33}^+ - f_{33}^{3+}~,  \label{noiso_reac6}\\ 
 f_{\pi^+ n \to \pi^0 p} & = & -\frac{1}{\sqrt{2}} \Big(f_{13}^- + f_{13}^{3-} \Big) \no\\ 
                         & = & f_{\pi^0 p \to \pi^+ n}~, \label{noiso_reac7}\\ 
 f_{\pi^- p \to \pi^0 n} & = & -\frac{1}{\sqrt{2}} \Big(f_{13}^- - f_{13}^{3-} \Big) \no\\ 
                         & = & f_{\pi^0 n \to \pi^- p}~, \label{noiso_reac8} 
\eea 
where, for simplicity of notation, we do not display the angular indices. 
Here we made use of the fact that  
$T_{11}^{+/3+}=T_{22}^{+/3+}$,  
$T_{12}^{-/3-}=T_{21}^{-/3-}$, 
$T_{13}^-=T_{23}^-=T_{31}^-=T_{32}^-$, and 
$T_{13}^{3-}=T_{23}^{3-}=-T_{31}^{3-}=-T_{32}^{3-}$. 
In eqs.~(\ref{noiso_reac7}) and (\ref{noiso_reac8}), we have also used time reversal invariance.  
 
\subsection{Isospin relations 
\label{sec:noiso_isorel}} 
 
Isospin violation is best characterized in terms of 
quantities which are exactly zero in the isospin limit of equal quark 
masses and vanishing electromagnetic coupling.  
We have eight reaction channels, see eqs.~(\ref{noiso_reac1})--(\ref{noiso_reac8}), 
which in the case of isospin symmetry are entirely described 
in terms of two amplitudes. One can thus write down six  
independent isospin relations (see also~\cite{kg} for 
a general analysis) 
\bea  
R_1 & = &  
2 \, \frac{f_{\pi^+ p \to \pi^+ p} + f_{\pi^- p \to \pi^- p}  
- 2 \, f_{\pi^0 p \to \pi^0 p}}  
          {f_{\pi^+ p \to \pi^+ p} + f_{\pi^- p \to \pi^- p} + 2 \, 
            f_{\pi^0 p \to \pi^0 p}}  
\no \\ &=&   
2 \, \frac{f^+_{11} - f^+_{33}+f^{3+}_{11}- f^{3+}_{33}}  
          {f^+_{11} + f^+_{33}+f^{3+}_{11}+ f^{3+}_{33}}~,  
\label{IRel1} \\ 
R_2 & = &  
2 \, \frac{f_{\pi^+ p \to \pi^+ p} - f_{\pi^- p \to \pi^- p} - \sqrt{2} \, f_{\pi^- p \to \pi^0 n}}  
          {f_{\pi^+ p \to \pi^+ p} - f_{\pi^- p \to \pi^- p} + \sqrt{2} \, f_{\pi^- p \to \pi^0 n}}  
\no \\ &=&  
2 \,\frac{f^-_{12}-f^-_{13}+f^{3-}_{12}-f^{3-}_{13}}  
         {f^-_{12}+f^-_{13}+f^{3-}_{12}+f^{3-}_{13}}~,  
\label{IRel2} \\   
R_3 & = &  
2 \, \frac{f_{\pi^0 p \to \pi^+ n} - f_{\pi^- p \to \pi^0 n}}  
          {f_{\pi^0 p \to \pi^+ n} + f_{\pi^- p \to \pi^0 n}}  
= 
2 \, \frac{f^{3-}_{13}}{f^-_{13}}~,  
\label{IRel3} \\   
R_4 & = &  
2 \, \frac{f_{\pi^+ p \to \pi^+ p} - f_{\pi^- n \to \pi^- n}}  
          {f_{\pi^+ p \to \pi^+ p} + f_{\pi^- n \to \pi^- n}}  
=   
2 \, \frac{f^{3+}_{11}-f^{3-}_{12}}{f^+_{11}-f^-_{12}}~,  
\label{IRel4} \\ 
R_5 & = &  
2 \, \frac{f_{\pi^- p \to \pi^- p} - f_{\pi^+ n \to \pi^+ n}}  
          {f_{\pi^- p \to \pi^- p} + f_{\pi^+ n \to \pi^+ n}}  
=  
2 \, \frac{f^{3+}_{11}+f^{3-}_{12}}{f^+_{11}+f^-_{12}}~,  
\label{IRel5} \\ 
R_6 & = &  
2 \, \frac{f_{\pi^0 p \to \pi^0 p} - f_{\pi^0 n \to \pi^0 n}}  
          {f_{\pi^0 p \to \pi^0 p} + f_{\pi^0 n \to \pi^0 n}}  
 =   
2 \, \frac{f^{3+}_{33}}  
          {f^+_{33}}~.  
\label{IRel6}  
\eea  
Note that this is just one of many possibilities to define these 
isospin ratios, but we find it particularly suitable. 
Again, we do not display angular variables. In the following, we will concentrate 
on the S--wave ($R_{i,0+}$) and two P--waves ($R_{i,1\pm}$). 
The first two ratios, the so--called triangle relations, are based on the observation that in the 
isospin conserving case, the elastic scattering channels involving charged 
pions are trivially linked to the corresponding neutral--pion elastic scattering or 
the corresponding charge exchange amplitude. 
To be precise, these ratios are to be formed with the real parts of 
the corresponding amplitudes. The imaginary parts of some of the amplitudes 
will be discussed later.  
Of particular interest is the second ratio, which is often  
referred to as {\it the} triangle relation. Only in this case all three 
channels have been measured (for pion kinetic energies as low as 30~MeV in the 
center--of--mass system) and the 7\% strong isospin violation reported  
in~\cite{gibbs,matsi} 
refer to this ratio. We stress again 
that it is difficult to compare this number to the one we will obtain in 
our calculation since a very different method of separating the electromagnetic 
effects is used. The ratio $R_6$ parameterizes 
the large isospin violation effect for $\pi^0$ scattering off nucleons 
first found by Weinberg~\cite{wein} and 
sharpened in~\cite{ms}, $R_6 \simeq 25\%$ at threshold. Note that in $R_1$, just as in 
$R_6$, the isovector terms drop out completely and one thus expects also a large 
isospin violation in this ratio (since the isoscalar parts are strongly 
suppressed and are of the same size as  the symmetry breaking terms). 
To our knowledge, this particular ratio 
has been called attention to for the first time in~\cite{fmsi}.  
{}From an experimental point of view, it 
has the advantage of avoiding the almost unmeasurable $n\pi^0$ 
amplitude appearing in $R_6$. On the other hand, as pointed out in ref.~\cite{aronprop}, the 
elusive $\pi^0 p$ scattering length might be measurable in neutral--pion  
photoproduction off protons below the $\pi^+ n$ threshold. This
results from the fact that the
imaginary part of the electric dipole amplitude $E_{0+}$ is proportional to the
respective $\pi N$ scattering length (final--state theorem); below the
$\pi^+ n$ threshold, one is thus exclusively sensitive to the small
$\pi^0 p$ phase shift.
Since the two ratios $R_1$ and $R_6$ are
entirely given in terms of isoscalar quantities, we expect large isospin  
symmetry breaking effects. But on the other hand, this also makes a precise determination 
of these ratios very difficult; indeed, the S--wave version of both these ratios  
is sensitive to the precise value of the combination of  
dimension--two LECs $2c_1-c_2-c_3$,  
but the strong contribution to the isoscalar scattering length is not even known in sign at present. 
The predictions for the other ratios are more stable since they 
involve the larger (and better determined) isovector quantities. 
In what follows, we will 
calculate the six ratios $R_i$ to leading one--loop accuracy, i.e.\ to third order 
in small momenta. For that, we have to consider tree graphs, some with 
fixed coefficients and some with LECs, and the leading one--loop graphs 
involving lowest order couplings only. 
 
\section{Calculation of the pion--nucleon amplitude} 
\label{sec:noiso_amplitude} 
 
As already mentioned, we will work to third order in the framework of 
heavy--baryon chiral perturbation theory. Since this method is well 
documented in the literature, see e.g.\ the comprehensive 
review~\cite{BKMrev}, we only discuss those terms in the isospin violating  
effective Lagrangian which are relevant for our analysis. For the 
particular case of isospin symmetric $\pi N$ scattering, a detailed 
exposition can be found in~\cite{FMS,pin4}. 
 
\medskip\noindent 
In the purely mesonic sector, we have to take into account the following structures: 
\be 
\label{Lpipi} 
{\cal L}_{\gamma^* \pi \pi}^{(2)} =  
\frac{F^2}{4} \lanl u_\mu u^\mu +\chi_+ \ranl + C \lanl Q_+^2 - Q_-^2 \ranl~, 
\end{equation} 
with  
\be 
Q=\frac{e}{2} (1+\tau_3)  
\end{equation} 
the nucleon charge matrix and $Q_\pm$ defined as follows: 
\be 
Q_{\pm} = \frac{1}{2} (u^\dagger Q u \pm u Q u^\dagger)~. 
\end{equation} 
$F$ is the pion decay constant in the chiral limit and the 
chiral vielbein is given by $u_\mu = i u^\dagger \del_\mu U u^\dagger$. 
The scalar source $\chi$ includes the quark mass matrix,  
$\chi = 2 B_0 {\cal M}$, with $B_0$ being proportional to the  
vacuum expectation value of the scalar quark density. 
We work in the standard scenario of chiral symmetry breaking, $B_0 \gg F$. 
{}From $\chi$ one constructs the fields  
$\chi_\pm = u^\dagger \chi u^\dagger \pm u \chi^\dagger u$. 
Furthermore, $\lanl \ldots \ranl$  denotes the trace in isospin space. 
For the pion masses to leading order this gives: 
\bea 
M_{0}^2 & = & B_0 (m_u + m_d)~, \no\\ 
M_{+}^2 & = & B_0 (m_u + m_d) + \frac{2}{F^2}  e^2 C~. 
\label{pionmassdiff} 
\eea 
The experimentally known pion mass difference thus directly fixes the 
value of the low--energy constant $C$, $C = 5.9\cdot 10^{-5}$ GeV$^4$.  
Note that the pion mass difference up to second chiral order is of  
electromagnetic origin only. 
 
\medskip\noindent 
In the one--nucleon sector, we take into account all terms related 
to strong isospin breaking as well as those electromagnetic terms, 
which contribute to the nucleon masses. The relevant structures in the 
Lagrangian are~\cite{svenphd} 
\bea 
\label{Lpin} 
&&{\cal L}_{\gamma^* \pi N} =\bar{N} \Big[  i v\cdot D + c_1 \lanl 
\chi_+  
\ranl +c_5 \widetilde{\chi}_+ \no\\&& 
+ F^2\Big(f_1 \langle Q_+^2 - Q_-^2 \rangle + f_2 \langle Q_+ \ranl 
Q_+  
+ f_3 \lanl Q_+ \ranl ^2 \Big)\no\\&& 
+d_{17} \lanl S\cdot u \chi_+ \ranl + i d_{18} [S\cdot D, \chi_-]  
+ i d_{19} [S \cdot D, \lanl \chi_- \ranl ] \no\\&& 
+ \ldots \Big] N~. 
\eea 
Here, $D_\mu$ is the covariant derivative, $S_\mu$ the nucleon spin 
vector and the $\widetilde{\chi}$ denotes the traceless part of $\chi$ (for 
further definitions and the remaining isospin symmetric structures,  
see~\cite{FMS}). For the masses up to  
${\cal O}(q^2)$ this means 
\bea 
m_N &=& m -4 M_0^2 c_1 -2 B_0 (m_u - m_d) c_5 \tau_3 \no\\ 
&& - e^2 F^2 [f_1 + \frac{f_2}{2} (1+\tau_3)+ f_3 ]~. 
\label{nucmassdiff} 
\eea 
The constant shift $-4 M_0^2 c_1- e^2 F^2 (f_1 + f_2/2+ f_3 )$ is unobservable. The  
proton--to--neutron mass difference fixes the values of the LECs through 
\bea 
(m_p- m_n)^{\rm str.} & = & -4 B_0 (m_u - m_d) c_5 \nonumber \\ 
&=& (-2.05 \pm 0.30)~{\rm MeV}~,\label{diffem}\\ 
(m_p- m_n)^{\rm em.} & = & - e^2 F^2 f_2 = (0.76\mp 0.30)~{\rm MeV}~.\label{diffst} 
\eea 
These values are taken from~\cite{GL82}, where the electromagnetic mass 
shift was calculated with the help of the Cottingham formula.  
Eq.~(\ref{diffst}) directly fixes the value of the electromagnetic 
coupling $f_2$, $f_2 = (-0.45 \pm 0.19)~{\rm GeV}^{-1}$. 
We then fix $B_0 (m_u - m_d) c_5$ in such a way that the 
neutron--proton mass difference coincides with the experimentally 
determined value. 
We want to stress that, contrary to the pion case, the nucleon mass difference is 
of electromagnetic {\it and} strong nature. 
Besides $c_1$, also $c_5, f_1$, and $f_2$ contribute to the $\pi \pi N N$ vertex 
which is given by 
\bea 
\label{pipinn} 
&& -i \frac{4}{F^2} c_1 M_0^2 \delta^{ab} 
-i c_5 B_0 (m_u-m_d) \frac{1}{F^2} \no\\ 
&&\hspace{5mm} \times \Big[ 2 \delta^{ab} \delta^{a3}   
     + i \epsilon^{bac} \tau^c  (\delta^{a3}-\delta^{b3} ) \Big] \tau^3\no\\ 
&&-i e^2 \Big[ 2 f_1 \delta^{ab} (1-\delta^{a3}) +\frac{f_2}{4} (2 
\delta^{ab} \tau^3 (1-\delta^{a3}) \no\\ 
&&\hspace{5mm}+i \epsilon^{bac} \tau^c \tau^3 (\delta^{b3}-\delta^{a3}) )\Big]~. 
\eea 
Whereas $c_5$ and $f_2$ can be fixed through eqs.~(\ref{diffem}) and (\ref{diffst})  
respectively, the value of $f_1$ is unknown. 
In our analysis, it will be varied in some reasonable range. 
Let us estimate these bounds by observing that the $f_1$--contribution to the nucleon mass shift  
should be of the same order of magnitude as any photonic loop, 
thus  
\begin{equation} 
e^2 F_\pi^2 f_1 \sim \left(\frac{e}{4 \pi} \right)^2 m_N~\,\, , 
\end{equation} 
which leads to $f_1 \sim 1$~GeV$^{-1}$ as a natural order\footnote{ 
  This dimensional analysis differs from the one of~\cite{MM}, 
  which considers $f_1 \sim 1/(4 \pi)$~GeV$^{-1}$ as a natural order of magnitude.}. 
In the following, we will use the range $e^2 f_1 = 0.0\pm 0.1~{\rm GeV}^{-1}$.  
We will also need the values of the counterterms related to isospin symmetric 
structures. The LECs $c_i$ and $d_i$ are taken from fit~1, 2, 3  
of~\cite{FMS}. These are fits of the third order isospin 
symmetric amplitude to the low--energy pion--nucleon 
phase shift analyses of refs.~\cite{koch}, \cite{mats}, and~\cite{SAID}, respectively. 
 
\medskip\noindent 
In terms of the operators defined in eqs.~(\ref{Lpipi}) and (\ref{Lpin}), retaining only the 
terms leading to the strong and electromagnetic hadron masses 
is achieved by setting 
\beq 
Ce^2 \neq 0\, ,\,f_i e^2 \neq 0\,  , \, m_u - m_d \neq 0\, , \quad 
{\rm but }\quad e^2 = 0~. 
\end{equation}  
 
\begin{figure}[h]
   \begin{center}
   \epsfxsize=5.3cm
   \epsffile{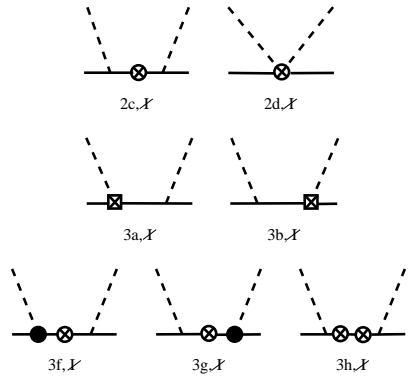}
   \vspace{0.5cm}
   \caption{\label{fig:noiso_tree}
   Tree graphs contributing to isospin violation in  $\pi N$ 
   scattering. The circle--cross (box--cross) refers to a dimension--two
   (--three) isospin violating vertex. The full circle stands for an
   isospin symmetric vertex of second order.
}
\end{center}
\end{figure}

\medskip\noindent 
The isospin symmetric tree graphs calculated to first, second, and 
third order, as well as the loop diagrams can be found in~\cite{moj}. 
The additional tree graphs with explicit isospin violating vertices 
($\sim f_1, f_2, \sim c_5, \sim \bar{d}_{17}, \bar{d}_{18}, \bar{d}_{19}$)  
are shown in fig.~\ref{fig:noiso_tree}.  
Before giving results, some important 
remarks concerning the chiral power counting are in order. Although   
the so--called Weinberg--Tomozawa $\bar{N}N\pi\pi$ contact graph gives 
a first order contribution to $g_{ba}^-$, in the ratios $R_i$ its 
effect is always proportional to  $\Delta \omega$, which is of second order, 
see eq.~(\ref{Dom}). 
Consequently, isospin violation only starts at second order in the chiral 
expansion. The chiral power counting is not as straightforward in the 
case of isospin violation as  in the isospin symmetric case. 
The difference of the pion energies is of second order, and should 
thus be neglected whenever it only leads 
to contributions of ${\cal O}(q^4)$. However, if this was done systematically, 
a diagram, which in fact vanishes at threshold, would now 
give a finite contribution. This can be easily seen in the following example:  
At threshold, either the incoming or the outgoing pion are at rest,  
and thus $\vec{q}_1 \cdot \vec{q}_2 = 0$. In a strict chiral expansion,  
this should be replaced by $(2 \w^2 - M_a^2 - M_b^2 -t)/2$, 
which does not vanish for $M_a \ne M_b$. In order to avoid such spurious 
effects, we decided also to account for $\Delta \w$ in some cases 
where it leads to a higher order effect, giving priority to the correct  
description of the amplitude at threshold. This is of course fully legitimate,  
since we only intend to give the amplitude correctly up to third order in 
the small expansion parameter $q$.

\section{Results} 
\label{sec:noiso_results} 
 
In this section, we want to generalize the analysis of ref.~\cite{fmsi} 
to the energy range above threshold. In addition, we give some additional
threshold results not shown in that paper.

\subsection{Momentum dependence} 
\label{sec:noiso_mom} 
 
Above threshold, we need the spin non--flip amplitudes  $g^{\pm, 3\pm}$ and 
the spin--flip ones, 
$h^{\pm, 3\pm}$. The expressions for the amplitudes are given  
in~\cite{nadiaphd}. 
We will focus on the results from fit~1 (based on the KA85 phase shifts). 
This analysis is different from the one in ref.~\cite{fmsi} in one 
important point: In~\cite{fmsi}, different reactions were compared 
at their particular threshold energy; this means that the ratios were formed  
by combining amplitudes at different center--of--mass energies. This will not  
be the case anymore. 
We will always consider $\sqrt{s}$ as a fixed quantity when forming ratios. 
This explains the difference between the numbers at low energy  
obtained here and the ones presented in~\cite{fmsi}. 
 
\begin{figure}[h]
\begin{center}   
   \epsfysize=3.2in
   \epsffile{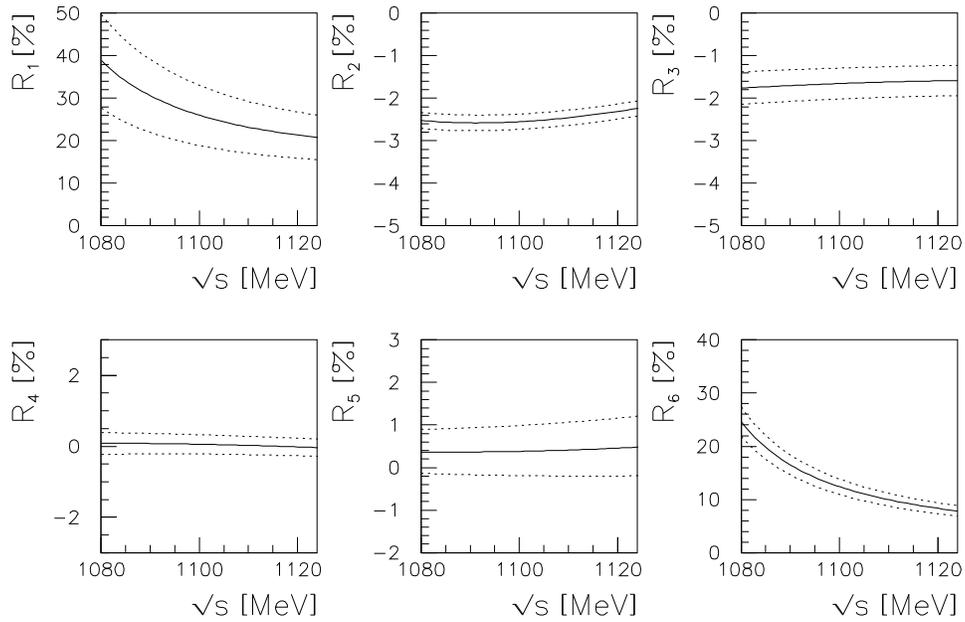}
   \vspace{1.5cm}
   \caption{\label{fig:noiso_ka85S}
   Isospin violation in the S--wave projection of
   the ratios $R_1$ -- $R_6$. The dashed lines indicate the range
   for isospin violation if the parameters $f_1$ and $f_2$ are varied
   as discussed in the text.
}
\end{center}
\end{figure}

\medskip\noindent 
In fig.~\ref{fig:noiso_ka85S}, we show the isospin breaking 
effects on the S--waves for the different ratios $R_1$ -- $R_6$, as a function of 
the total center--of--mass energy $\sqrt{s}$. 
Note that in these calculations, the spinor normalization factors 
are also taken into account\footnote{This is in contrast to 
  what was done in~\cite{fmsi}, but the inclusion of these factors  
  is necessary if one wants to define 
  the partial waves in the standard way. We have checked that the conclusions  
  concerning isospin violation do not depend on whether we include this factor  
  or not.}.  
As expected, the purely isoscalar ratios $R_1$ and $R_6$  
show very large isospin breaking effects. Those ratios, on 
the other hand, which are given in terms of both isoscalar and isovector 
quantities show a rather small effect, in the order of 
1--2\%. 
Also, the energy dependence is as expected: the lower the energy,  
the more important the quark mass difference is compared to 
the kinetic energies of the particles, and the larger the 
isospin breaking effects should be. As well for $R_1$ as for $R_6$, 
the effect drops by a factor of two over the first 50 MeV.  
The ratios $R_2$ -- $R_5$ remain almost constant over 
the whole energy range displayed here. The previously discussed
theoretical uncertainties due to the lack of knowledge of $f_1$ and $f_2$ 
lead to the bands displayed  
in fig.~\ref{fig:noiso_ka85S}\footnote{We stress  
  again that a variation in $f_2$ stands for 
  varying $f_2$ and $4B_0 (m_u-m_d)c_5$ such that the neutron--proton mass 
  difference remains constant.}.  
It is important to note that the variation in $f_1$ only shows up in  
$R_1$, $R_4$, and $R_5$.  
Whereas the range in $f_1$ dominates the uncertainty in $R_1$, 
it leads to vanishingly small contributions to $R_4$ and $R_5$. 
It is thus the uncertainty in $f_2$ which largely dominates the width of the 
bands in $R_2$ -- $R_6$. Since electromagnetic effects proportional
to $f_1$ and $f_2$ 
do not contribute to neutral--pion scattering, the band for $R_6$  
is entirely due to the variation in $B_0(m_u-m_d)c_5$. 
We see that the prediction for $R_1$ is quite sensitive to the precise value of the  
parameter $f_1$. But the conclusions remain unchanged:  
we expect huge isospin violation in the S--wave parts of the 
purely isoscalar ratios $R_1$ and $R_6$.

\begin{figure}[h]
   \begin{center}
   \epsfysize=3.2in
   \epsffile{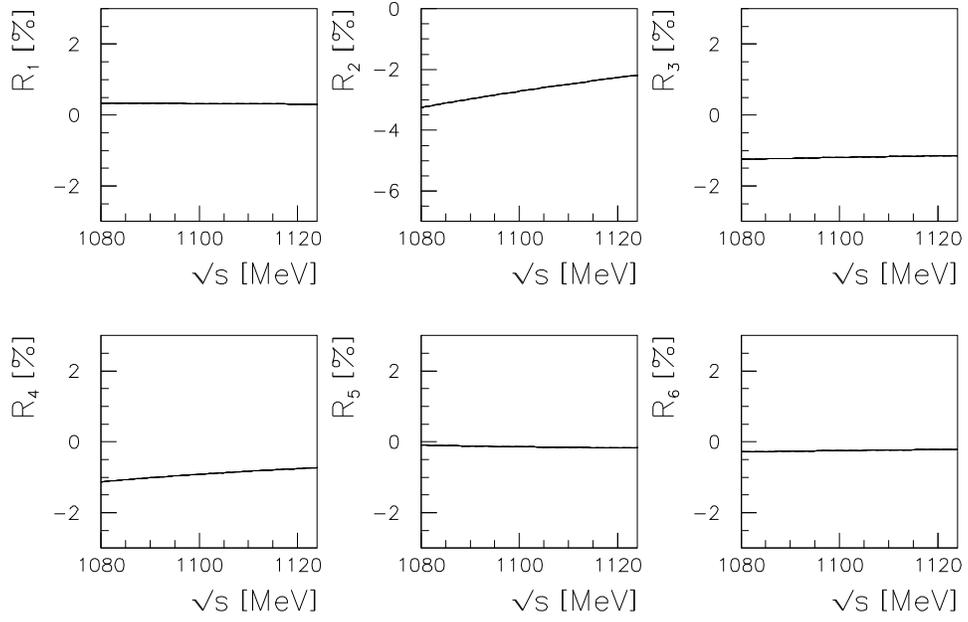}
   \vspace{1.5cm}
   \caption{\label{fig:noiso_ka85P3}
   Isospin violation in the ${\rm P}_3$--wave projection of
   the ratios $R_1$ -- $R_6$.
}
\end{center}
\end{figure}

\begin{figure}[p]
   \begin{center}
   \vspace{1.5cm}
   \epsfysize=3.2in
   \epsffile{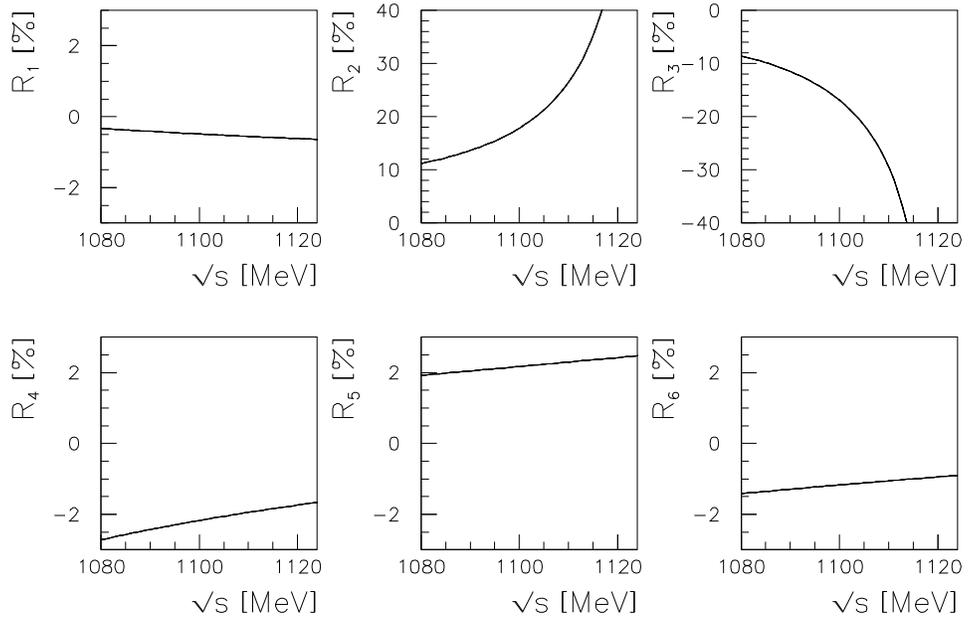}
   \vspace{1.5cm}
   \caption{\label{fig:noiso_ka85P1}
   Isospin violation in the ${\rm P}_1$--wave projection of
   the ratios $R_1$ -- $R_6$.
}   
\end{center}
\end{figure}

\medskip\noindent 
The corresponding P$_3$-- and P$_1$--wave projections of these relations are shown in 
figs.~\ref{fig:noiso_ka85P3} and \ref{fig:noiso_ka85P1}.  
The notation is standard: $l=1$ for both P--waves, for {${\rm P}_3$} the total angular momentum is 
$3/2$ whereas {${\rm P}_1$} is related to $j=1/2$. For ${\rm P}_3$, everything is as expected:  
isospin breaking is negligibly small, as well in the purely isoscalar channels 
as in the ones given by isoscalar and isovector quantities. It is interesting 
to note that in this projection, $R_2$ turns out to show the biggest  
effect, about $-3\%$ close to threshold. 
However, for the ${\rm P}_1$--waves, the results are more surprising:  
indeed for $R_2$ and for $R_3$, 
the effects from isospin violation become larger with increasing center--of--mass 
energy: 30~MeV above threshold, they are as big as $30\%$ for $R_2$ and $R_3$. 
But we run into the same problems as in the threshold analysis~\cite{fmsi}, 
namely that the denominators of some of the relations 
become very small in the ${\rm P}_1$--projection. We find a really spectacular result indicating 
isospin violation as big as $30\%$ in $R_2$ and $R_3$, but it will be a hopeless case 
to measure such effects, since the associated isospin conserving part is vanishingly
small as well. 
It is thus necessary to find a different representation, to project out 
different quantities than the usual partial waves. 
From~(\ref{partwavv}) we obtain 
\bea 
{\rm P}_3 (s) &=& {\cal G}(s) + {\cal H}(s)~, \no\\ 
{\rm P}_1 (s) &=& {\cal G}(s) -2 {\cal H}(s)~,  
\eea 
with 
\bea 
{\cal G}(s) & = & {{\sqrt{E_1+m_1} \sqrt{E_2+m_2}} \over 16 \pi \sqrt{s}}  
\int_{-1}^{+1} dz \, g_{ba} \, P_1 (z) ~,\no\\ 
{\cal H}(s) & = & {{\sqrt{E_1+m_1} \sqrt{E_2+m_2}} \over 16 \pi \sqrt{s}}  
\int_{-1}^{+1} dz \, \vec{q}_1\cdot \vec{q}_2 \, h_{ba}\no \\  
&& \quad \times\big(P_2(z) -z P_1 (z) \big) ~. 
\eea 
In figs.~\ref{fig:noiso_ka85g} and \ref{fig:noiso_ka85h}, we thus show  
isospin breaking in the ratios $R_1$ -- $R_6$ formed with the quantities 
${\cal G}$ and ${\cal H}$, respectively. 
{}From these figures we can see that indeed there is no large isospin symmetry 
breaking in the P--waves. The problems for the representation 
in terms of P$_1$ and P$_3$ only resulted from very small 
denominators. Interestingly, the ratios $R_1$ and $R_6$, 
which were found to give large results in the S--wave 
projection, are now found to be very small. Their ${\cal G}$-- and  
${\cal H}$--projection show an effect much smaller than $1\%$. It seems 
that for the P--wave projection, $R_2$ is the most promising ratio 
to study. Especially the ${\cal H}$--projection is very large and 
indicates isospin breaking of $-6\%$ at threshold. Note that neither 
the variation of $f_1$ nor the combined one of 
$f_2$ and $c_5$ affects the P--wave projections of our ratios. 
 
\begin{figure}[hbt] 
   \begin{center} 
   \epsfysize=2.4in 
   \epsffile{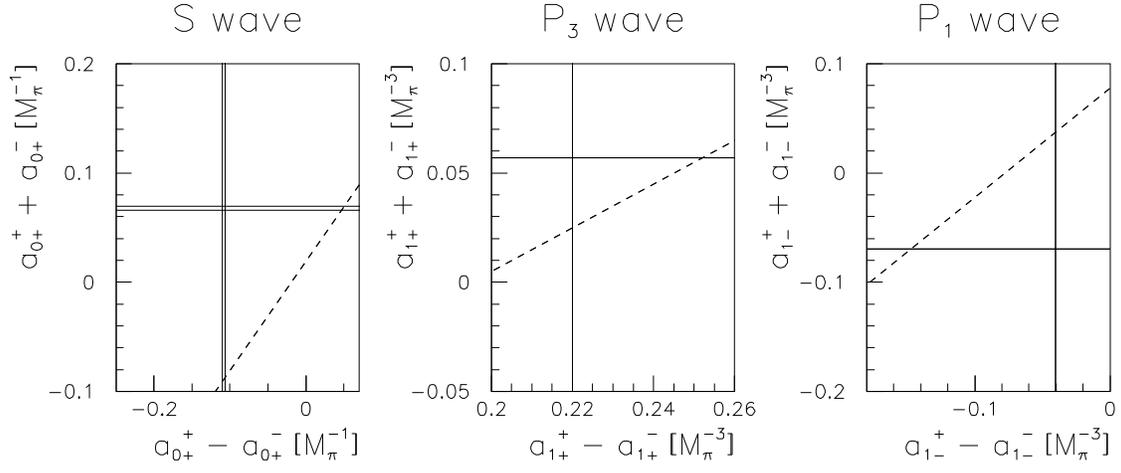} 
   \vspace{1.5cm} 
   \caption{\label{fig:noiso_em} 
   Real parts of the scattering lengths and volumes. The $\pi^+ p$ elastic process  
   is represented by the vertical bands, the $\pi^- p$ elastic process by the horizontal 
   bands, and the charge exchange reaction $\pi^- p \to \pi^0 n$ by the 
   diagonal bands.} 
\end{center} 
\end{figure}

\medskip\noindent 
In fig.~\ref{fig:noiso_em}, we show an analogue representation of 
isospin breaking effects to what was presented in~\cite{matsi}. 
The plot is based on the philosophy put forward in~\cite{BaL}. Provided that 
isospin is a good symmetry, the entire information concerning the elastic 
$\pi^\pm p$ and the charge exchange $\pi^- p \to \pi^0 n$ reactions is contained in 
two energy--dependent complex functions per partial wave, $f^+$ and $f^-$.  
In such a case, the $\pi^+ p$ reaction is described by $f^+ - f^-$ (represented by the  
vertical band in fig.~\ref{fig:noiso_em}), the $\pi^- p$ reaction by 
$f^+ + f^-$ (horizontal band in fig.~\ref{fig:noiso_em}), and the charge exchange  
reaction is given by $-\sqrt{2} f^-$ (shown by the diagonal band in fig.~\ref{fig:noiso_em}). 
At any energy, the principle of isospin symmetry necessitates a common overlap 
of the three bands. Here, we only show the corresponding bands at threshold. Note that  
most of the bands are degenerate to a line, since we have not 
taken into account any experimental errors, but only the spread due to the 
lack of knowledge of the value of $f_1$. 
In this way of looking at the problem, we cannot quantify isospin 
violation precisely, since we do not have any experimental errors, which would allow us, 
as in the case of~\cite{matsi}, to give the standard deviation of the difference 
between the charge exchange band and the intersection of the two elastic bands.

\subsection{Threshold analysis 
\label{sec:noiso_thresh}} 
 
The threshold region was already discussed in ref.~\cite{fmsi}, in 
particular the large effect on the ratio $R_1$ was for the first 
time pointed out in that paper and the influence of the operator $\sim c_5$ was 
analyzed. In that paper, predictions for pion scattering off protons 
were given. For completeness, in table~\ref{tab:a}, we collect these 
predictions together with the ones for scattering off neutrons (for 
the central values of the LECs $c_5$, $f_1$, and $f_2$). 
The large difference among the three sets of predicted values  
comes from the various input data; 
in particular, the theoretically most interesting scattering 
lengths for neutral--pion scattering off nucleons cannot be predicted 
very reliably. This theoretical uncertainty calls for a dedicated pion 
photoproduction experiment $\gamma p \to \pi^0 p$ below the 
secondary threshold which will allow to pin down the scattering length $a(\pi^0 p)$,  
as detailed in ref.~\cite{aronprop}. For a more precise discussion 
on isospin violation effects in threshold S--waves, see~\cite{fmsi}. 
\renewcommand{\arraystretch}{1.2} 
\begin{table}[htb] 
\begin{center} 
\begin{tabular}{|c||c|c|c|} 
   & Fit 1 & Fit 2 & Fit 3 \\ 
    \hline\hline 
$a(\pi^+ p \to \pi^+ p)$& $-108.8$         & $-83.8$          & $-95.3$ \\ 
$a(\pi^- p \to \pi^- p)$& $70.5 + i\,3.7 $ & $71.3 + i\,3.7 $ & $76.9 + i\,3.7 $ \\ 
$a(\pi^0 p \to \pi^0 p)$& $-13.4$          & $-0.1$           & $-2.6$\\ 
$a(\pi^+ n \to \pi^+ n)$& $69.7 + i\,3.7 $ & $70.5 + i\,3.7 $ & $76.2 + i\,3.7$ \\ 
$a(\pi^- n \to \pi^- n)$& $-109.6$         & $-84.6$          & $-96.1$ \\ 
$a(\pi^0 n \to \pi^0 n)$& $-11.0$          & $ 2.2$           & $-0.2$\\ 
$a(\pi^- p \to \pi^0 n)$& $-125.7$         & $-108.5$         & $-120.8$\\ 
$a(\pi^0 p \to \pi^+ n)$& $-124.9 -i\,0.6 $& $-107.8 -i\,0.6 $& $-119.9 -i\,0.6$\\ 
  \end{tabular} 
  \vspace{0.2cm} 
  \caption{Values of the scattering lengths for pion scattering off 
   nucleons in units of $10^{-3}/M_{\pi^+}$ 
   for the various parameter sets as given by the 
   fits~1, 2, 3 of ref.~\protect\cite{FMS}.\label{tab:a}} 
\end{center}\end{table} 
 
\subsection{Comparison to other analyses} 
\label{sec:noiso_comp} 
 
The question of isospin violation in pion--nucleon scattering has stimulated vast 
efforts on the experimental, but also on the phenomenological side. 
 
\medskip\noindent 
Gibbs, Li Ai, and Kaufmann~\cite{gibbs} restrict their analysis to the triangle relation 
(arguing that only in this channel, all reactions are experimentally accessible) 
and look for deviations from zero of  
$D = f_{\rm CEX} - (f_{\pi^+ p} - f_{\pi^- p})/\sqrt{2}$. 
They are only interested in isospin violation which lies beyond 
effects due to the hadronic mass differences and the Coulomb corrections. 
They thus describe different physics than we do, and our results should consequently 
not be directly compared to theirs. They find $D=(-0.012\pm 0.003)~{\rm fm}$, thus a four--$\sigma$--effect. 
They cannot conclude where the breaking actually occurs; it could as well be in the 
charge exchange, in the elastic amplitudes, or in a combination of both. 
 
\medskip\noindent 
Matsinos~\cite{matsi} analyzes the whole set of low--energy pion--nucleon scattering data 
with the help of an extended tree level model~\cite{leisi}. The known electromagnetic 
corrections and hadronic mass effects are taken care of by the NORDITA method~\cite{NO}. 
He finds consistency of the low--energy pion--nucleon database and 
comes to similar conclusions than Gibbs et al.: the relative difference in the 
real parts of the two S--wave amplitudes (the one of the elastic channels 
and the one corresponding to charge exchange) amounts to $(6.4\pm 1.4) \%$. 
 
\medskip\noindent 
Both authors are only interested in isospin violation stemming 
from strong vertices, neither in electromagnetic corrections 
to the amplitudes, nor in mass difference effects. This approach is  
completely different from ours and the results should not be naively 
compared. Both analyses come to the same conclusions, although they are based 
on different models for describing the strong part of the 
$\pi N$ interaction and they use different algorithms 
for separating strong effects from electromagnetic and mass 
difference ones. But we stress again that in the framework we are using, a consistent 
separation of the electromagnetic and the strong effects is possible 
and to our knowledge this has not been achieved before.  
Only when a mapping of the method developed here on the commonly used procedures of 
separating electromagnetic and hadronic mass effects (such as the NORDITA 
method~\cite{NO}) has been performed, a sensible comparison with the numbers quoted  
in the literature will be possible. In order to achieve this, further work in 
chiral perturbation theory calculations will have to be done. 
 
\begin{figure}[htb] 
\begin{center} 
   \epsfxsize=3.5in 
   \epsffile{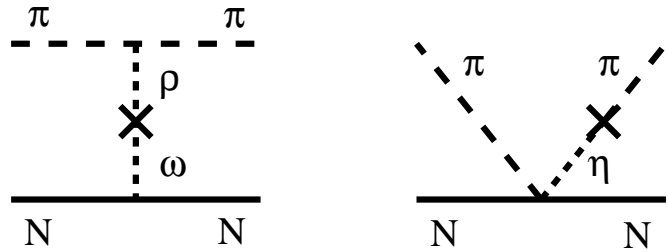} 
   \vspace{1.5cm} 
   \caption{\label{fig:noiso_xchmod} 
   Feynman diagrams commonly used in meson exchange models 
   to explain isospin violation in pion--nucleon scattering.} 
\end{center} 
\end{figure}

\medskip\noindent 
In meson exchange models, isospin violation in pion--nucleon scattering 
is mainly accounted for by $\pi\eta$ mixing. Pion--nucleon scattering then proceeds 
e.g.\ through an $\eta$ production process followed by an isospin symmetry  
breaking transition $\eta \to \pi^0$. This process can only occur  
if at least one of the pions is neutral.  
In an SU(2) approach to chiral perturbation theory, $\pi\eta$  
mixing is hidden in some counterterms. If one performs an SU(3) calculation,  
one can explicitly take into account such effects, and one can convince oneself that  
they are indeed proportional to $m_u-m_d$~\cite{GL82,IS80}. 
Another process which could explain isospin violation is 
$\rho\omega$ mixing~\cite{bg94}. Although this effect can account for the same 
magnitude of isospin breaking as observed in~\cite{gibbs} and \cite{matsi}, 
it gives the wrong sign. Both diagrams can be seen in fig.~\ref{fig:noiso_xchmod}. 
 
\section{Summary} 
\label{sec:noiso_summary} 
We have presented an analysis of isospin breaking effects in   
pion--nucleon scattering  due to the light quark mass difference and the  
dominant virtual--photon effects.  
\begin{enumerate} 
\item[(i)]  
We have taken into account all operators related to strong isospin breaking  
and the electromagnetic ones that lead to the pion and nucleon mass differences.  
Stated differently, the finite parts of some of the virtual--photon operators  
contributing at this order have been set to zero.  
This allows in particular to isolate the contribution of the strong  
isospin breaking dimension--two isovector operator first  
considered by Weinberg. We have considered 
a set of six ratios $R_i$, which vanish in the limit of isospin conservation. 
{}From these, four involve isovector {\it and} isoscalar amplitudes ($R_{2,3,4,5}$)  
while the two others are of purely isoscalar type $(R_{1,6})$.  
\item[(ii)]  
We have extended the analysis of ref.~\cite{fmsi} to higher center--of--mass energies. 
In the S--wave, isospin violating effects tend to disappear rather 
quickly in energy. But in the P$_1$--wave, due to a very small 
isospin symmetric part, relative isospin violation 
becomes very large in some ratios. In order to give a more reliable  
description of the phenomenon, we presented isospin breaking 
in two other quantities, which are more directly related to the spin--flip  
and spin--non--flip amplitudes. We conclude that isospin violation effects are 
small in these new projections. 
\item[(iii)]  
We have tabulated the theoretical predictions for S--wave scattering lengths 
in the eight physical channels and stressed the importance of 
measuring the elusive $\pi^0 p$ channel via precise photoproduction 
experiments (which should be feasible at MAMI or the TUNL--FELL). 
\end{enumerate} 
 
\medskip\noindent 
We want to note again that within the framework presented here, a unique and unambiguous separation 
of all different isospin violating effects is possible. To access the size of 
isospin violation encoded in the presently available pion--nucleon scattering data, 
the extension of this scheme to include hard and soft photons is mandatory. 
Once this is done, it will be possible to analyze the cross 
section data directly without  
recourse to any model for separating electromagnetic  
or hadronic mass effects, thus avoiding any 
mismatch by combining different approaches or models.

\bigskip\bigskip 
 
\section*{Acknowledgments} 
 
We would like to thank Aron Bernstein for very stimulating discussions.

\vspace{1cm} 

\vfill

\onecolumn 

\section*{Figures}

\vspace{1cm}

$\,$

\vspace{0.5cm}

\noindent
\begin{figure}[p]
   \begin{center}
   \epsfysize=3.2in
   \epsffile{pg32.epsi}
   \vspace{1.5cm}
   \caption{\label{fig:noiso_ka85g}
   Isospin violation in the ${\cal G}$ projection of
   the ratios $R_1$ -- $R_6$.
}
\end{center}
\end{figure}

\begin{figure}[p]
   \begin{center}
   \vspace{1.5cm}
   \epsfysize=3.2in
   \epsffile{ph32.epsi}
   \vspace{1.5cm}
   \caption{\label{fig:noiso_ka85h}
   Isospin violation in the ${\cal H}$ projection of
   the ratios $R_1$ -- $R_6$.
}
\end{center}
\end{figure}

\end{document}